\documentclass{article}
\usepackage{epsfig}
\usepackage{a4wide}
\newcommand{\GeV}{\mbox{ GeV}} 
\newcommand{\TeV}{\mbox{ TeV}}
 
\newcommand{\sq}{\ensuremath{\tilde q}} 

\newcommand{\stopp}{\ensuremath{\tilde t}} 
\newcommand{\cplus}{\ensuremath{\chi^+}} 
\newcommand{\neut}{\ensuremath{\chi^0}}

\newcommand{\sfr}{{\tilde{f}}}

\newcommand{\tb}{\ensuremath{\tan\beta}}

\newcommand{\mb}{\ensuremath{m_b}} 
\newcommand{\mt}{\ensuremath{m_t}}

\newcommand{\mw}{\ensuremath{M_W}} 
\newcommand{\mws}{\ensuremath{M^2_W}} 
 
\newcommand{\mxs}{\ensuremath{M^2_X}}
\newcommand{\mx}{\ensuremath{M_X}}

\newcommand{\mHp}{\ensuremath{M_{H^\pm}}} 

\newcommand{\mi}{\ensuremath{M_i}} 
\newcommand{\mis}{\ensuremath{M^2_i}}

\newcommand{\mg}{\ensuremath{m_{\tilde{g}}}} 
 
\newcommand{\msbo}{\ensuremath{m_{\tilde{b}_1}}} 
\newcommand{\msbt}{\ensuremath{m_{\tilde{b}_2}}} 
 
\newcommand{\msto}{\ensuremath{m_{\tilde{t}_1}}} 
\newcommand{\mstt}{\ensuremath{m_{\tilde{t}_2}}}

\newcommand{\msdo}{\ensuremath{m_{\tilde d_1}}} 
\newcommand{\msdt}{\ensuremath{m_{\tilde d_2}}} 
\newcommand{\msut}{\ensuremath{m_{\tilde u_2}}}

\newcommand{\mselo}{\ensuremath{m_{\tilde{e}_1}}} 
\newcommand{\mselt}{\ensuremath{m_{\tilde{e}_2}}} 
 
\newcommand{\mwf}{{\mw^4}}

\newcommand{\cbta}{c_\beta}
\newcommand{\cfbt}{c_{4\beta}}
\newcommand{\sbta}{s_\beta}
\newcommand{\stbt}{s_{2\beta}}

\newcommand{\sws}{s_W^2}

\newcommand{\UChaio}{U_{i1}}
\newcommand{\UChait}{U_{i2}}

\newcommand{\mchasomenysmchast}{\left(M_1^2-M_2^2\right)}
\newcommand{\commonnumfactor}{\frac{\alpha}{4\,\pi\,\sws}}

\newcommand{\osf}{\ensuremath{\theta_f}}
\newcommand{\osb}{\ensuremath{\theta_b}}
\newcommand{\ost}{\ensuremath{\theta_t}}
\newcommand{\osd}{\ensuremath{\theta_d}}
\newcommand{\osu}{\ensuremath{\theta_u}}
\newcommand{\osel}{\ensuremath{\theta_e}}

\begin{document}

\begin{flushright}
{\parbox{4cm}{PSI-PR-02-12\\
UB-ECM-PF-02-21\\
hep-ph/0210120 \\
October 2002
}}
\end{flushright}

\begin{center}

{\Large \textbf{Full one-loop corrections to sfermion
    decays}\footnote{Talk presented at the \textit{10th International Conference on Supersymmetry and Unification of
Fundamental Interactions} (SUSY 02), DESY, Hamburg, Germany, 17-23 June
    2002.}
} 

\vspace{0.8cm}

{\large \underline{Jaume Guasch}$^{a}$,  
Wolfgang Hollik$^{b}$, Joan Sol\`a$^{c,d}$}

\vspace*{0.2cm}
{\sl 
$^a$ Theory Group LTP, Paul
    Scherrer Institut, CH-5232 Villigen PSI, Switzerland\\
$^b$  Max-Planck-Institut f\"ur Physik,
   F\"ohringer Ring 6, D-80805 M\"unchen, Germany \\
$^c$  Departament d'Estructura i Constituents de la 
Mat\`eria,  Universitat de Barcelona, Diagonal 647, E-08028 Barcelona,
  Catalonia, Spain\\
$^d$ Institut de F\'{\i}sica d'Altes Energies, Universitat Aut\`onoma de
  Barcelona, E-08193 Bellaterra, Barcelona, Catalonia, Spain} 

\end{center}

 \begin{abstract}
\noindent
We analyze the partial decay widths of sfermions decaying into charginos
and neutralinos $\Gamma(\sfr\to f'\chi)$ at the one-loop level. We
present the renormalization framework, and discuss the value of the
corrections for top- and bottom-squark decays.
 \end{abstract}

\section{Introduction}

One of the basic predictions of Supersymmetry (SUSY) is the equality between
the couplings of SM particles and that of their superpartners. The
simplest processes in which this predicition could be tested, is the partial
decay widths of sfermions into Standard Model (SM) fermions and
charginos/neutralinos:
\begin{equation}
\Gamma(\sfr \to f' \chi)\,\,.
\label{eq:gammadef}
\end{equation}
By measuring these partial decay widths (or the corresponding branching
ratios) one could measure the fermion-sfermion-chargino/neutralino
Yukawa couplings and compare them with the SM fermion gauge couplings. 

We have computed the full one-loop electroweak corrections to the
partial decay widths~(\ref{eq:gammadef}). As we will show, the radiative
corrections induce finite shifts in the couplings which are non-decoupling.

The QCD corrections to the process~(\ref{eq:gammadef}) were computed
in~\cite{QCD}, 
and the Yukawa corrections to
bottom-squarks decaying into charginos was given in~\cite{Guasch:1998as}.
Here we present the last step, namely, the full electroweak corrections
in the framework of the Minimal Supersymmetric Standard Model (MSSM).
Full details of the present work can be found
in~\cite{EWcorr}. 

\section{Renormalization and radiative corrections}

The computation to one-loop level of the partial decay
width~(\ref{eq:gammadef}) requires the renormalization of the full MSSM
Lagrangian, taking into account the relations among the different
sectors and the mixing parameters. We choose to work in an on-shell
renormalization scheme, in which the renormalized parameters are the
measured quantities. 
The SM sector is renormalized according to the                                
standard on-shell SM $\alpha$-scheme~\cite{Hollik}, and the MSSM Higgs         
sector (in particular the renormalization of $\tb$) is treated as in~\cite{Dabels}.

As far as the sfermion sector is concerned, we follow the procedure
described in~\cite{Guasch:1998as}. However,  in the present
analysis we treat simultaneoulsy top-squarks and bottom-squarks. Due to
$SU(2)_L$ invariance the parameters in these two sectors are not
independent, and we can not supply with independent on-shell conditions
for both sectors. We choose as input parameters the on-shell masses of
both bottom-squarks, the lightest top-squark mass, and the mixing angles
in both sectors\footnote{Throughout this work we make use of third generation
notation. The notation is as in~\cite{Guasch:1998as,EWcorr}.}: 
\begin{equation}
  \label{eq:inputsf}
(m_{\tilde{b}_1}, m_{\tilde{b}_2}, \osb ,m_{\tilde{t}_2},
\ost), \ \ \ \ m_{\sfr_1}>m_{\sfr_2}.  
\end{equation}
 The remaining parameters are computed as a function of those
 in~(\ref{eq:inputsf}). In particular, the trilinear soft-SUSY-breaking
 couplings read:
\begin{equation}
A_{\{b,t\}}=\mu\{\tan\beta,\cot\beta\}+
{m_{\tilde{f}_1}^2-m_{\tilde{f}_2}^2\over 2\,m_f}\,\sin{2\,\osf}\,,
\label{eq:Abt}
\end{equation}
with $\tb=v_2/v_1$, the ratio of the vacuum expectation values of the
two Higgs boson doublets. The approximate (necessary) condition to avoid
colour-breaking minima in the MSSM Higgs potential~\cite{Frere:1983ag},
\begin{equation}
A_q^2<3\,(m_{\tilde{t}}^2+m_{\tilde{b}}^2+M_H^2+\mu^2)\,,
\label{eq:necessary}
\end{equation}
imposes a tight correlation between the sfermion mass splitting and the
mixing angle at large $\tb$.
 Since the heaviest top-squark mass ($\msto$)
is not an input 
parameter, it 
receives finite radiative corrections:
\begin{equation}
  \label{eq:mst1radcor}
  \Delta \msto^2 = \delta \msto^2 + \Sigma_{\stopp_1}(\msto^2) \,\,, 
\end{equation}
where $\delta\msto^2$ is a combination of the counterterms of the
parameters in~(\ref{eq:inputsf}), and the counterterms of the gauge and
Higgs sectors. 

The chargino/neutralino sector contains six particles, but only three
independent input parameters:   the soft-SUSY-breaking
$SU(2)_L$ and $U(1)_Y$ gaugino masses ($M$ and $M'$), and the higgsino mass parameter ($\mu$). The situation in this sector is quite
different from the 
sfermion case,
since in this case no independent
counterterms for the mixing matrix elements can be introduced. We stick
to the following procedure: First, we introduce a set of
\textit{renormalized} parameters $(M,M',\mu)$ in the expression of the
chargino and neutralino matrices ($\cal{M}$ and ${\cal M}^0$), and
diagonalize them  by means of unitary matrices $M_D=U^* {\cal
  M} V^\dagger$, $M_D^0 = N^* {\cal M}^0 N^{\dagger}$. Now $U$, $V$ and
$N$ must be regarded as \textit{renormalized mixing matrices}. The
counterterm mass matrices are then $\delta M_D=U^* \delta{\cal M}
V^\dagger$, $\delta M_D^0 = N^*\delta {\cal M}^0 N^{\dagger}$, which are
non-diagonal. At this point, we introduce renormalization conditions for
certain elements of $\delta M_D$ and $\delta M_D^0$. In particular, we use
on-shell renormalization conditions for the two chargino masses ($M_1$
and $M_2$), which allows to compute the counterterms $\delta M$ and
$\delta \mu$. This information, together with the on-shell condition for
the lightest neutralino mass ($M_1^0$) allows to derive the expression
for the counterterm $\delta M'$.
The other
neutralino masses ($M_{2,3,4}^0$) receive radiative corrections. In this
framework the 
renormalized one-loop 
chargino/neutralino 2-point functions 
are \textit{non}-diagonal. Therefore one must take into account this
mixing either by including explicitly the reducible $\chi_r-\chi_s$
mixing diagrams, or by means of  external mixing wave-function terms
(${\cal Z}_{\{L,R\}}^{0\beta \alpha}$, ${\cal Z}_{\{L,R\}}^{-ij}$).
See Refs.~\cite{EberlFritzsche} 
for different (but
one-loop equivalent) approaches to the renormalization of the
chargino/neutralino sector.\footnote{See Ref.~\cite{Majerotto} for a
  review of radiative corrections to SUSY processes.}

The complete one-loop computation consists of: 
\begin{itemize}
\item 
renormalization constants for the parameters and wave functions in
  the bare Lagrangian, 
\item
one-loop one-particle irreducible three-point functions, 
\item 
mixing terms among the external charginos and neutralinos, 
\item 
soft- and hard- photon  bremsstrahlung.
\end{itemize}
All kind of MSSM particles are taken into account in the loops: SM fermions,
sfermions, electroweak gauge bosons, Higgs bosons, Goldstone bosons,
Fadeev-Popov ghosts, charginos, neutralinos. The computation is
performed in the 't Hooft-Feynman gauge, using dimensional reduction
for the regularization of divergent integrals. 
The loop computation itself is done using the computer algebra packages
\textit{FeynArts 3.0} and \textit{FormCalc
  2.2}~\cite{FeynArts3,Hahn:1998yk}. 
The numerical evaluation of one-loop integrals makes
use of \textit{LoopTools 1.2}~\cite{Hahn:1998yk}.\footnote{The resulting
  FORTRAN code can be obtained from \texttt{http://www-itp.physik.uni-karlsruhe.de/$\sim$guasch/progs/}.}

\section{Results}

The results show the very interesting property that none of the particles
of the MSSM decouples from the corrections to the
observables~(\ref{eq:gammadef}). This can be well understood in terms of
renormalization group (RG) running of the parameters and SUSY
breaking. Take, e.g., the effects of squarks in the
electron-selectron-photino coupling. Above the squark mass scale
($Q>m_{\sq}$) the electron electromagnetic coupling ($\alpha(Q)$) is equal
(by SUSY) to the electron-selectron-photino coupling
($\tilde{\alpha}(Q)$), and both couplings run according to the same RG
equations. At $Q=m_{\sq}$ the squarks \textit{decouple} from the
RG running of the couplings. At $Q<m_{\sq}$, $\alpha(Q)$ runs due to the
contributions from pure quark loops, but $\tilde{\alpha}(Q)$ does not
run anymore, and it is
\textit{frozen} at the squark scale, that is:
$\tilde{\alpha}(Q<m_{\sq})=\alpha(m_{\sq})$.  Therefore, when comparing
these two couplings at a scale $Q<m_{\sq}$, they differ by  the
logarithmic running of $\alpha(Q)$ from the squark scale to $Q$:
$\tilde{\alpha}(Q)/\alpha(Q)-1=\beta \log(m_{\sq}/Q)$.

The above discussion has two important consequences: 
\begin{enumerate}
\item The
non-decoupling can be used to extract information of the high-energy
part of the SUSY spectrum: one can envisage a SUSY model in which a
significant splitting among the different SUSY masses exists,
e.g. $m_{\sq} \gg m_{\tilde l}$, where the sleptons lie below the
production threshold in an $e^+e^-$ linear collider, but the squarks are
above it. By means of high precision measurements of the
lepton-slepton-chargino/neutralino couplings one might be able to
extract information of the squark sector of the model, to be checked
with the available data from the LHC. 
\item  By the same token, it means
that the value of the radiative corrections depends on all
parameters of the model, and we can not make precise quantitative
statements unless the full SUSY spectrum is known. 
This drawback can be partially overcome by the introduction of                
\textit{effective coupling                                                     
matrices}, which can be defined as follows. The subset of fermion-sfermion     
one-loop contributions to                                                      
the self-energies of gauge-boson, Higgs-bosons, Goldstone-bosons, charginos    
and neutralinos form a gauge invariant finite subset of the corrections.       
Therefore these contributions can be absorbed into a finite shift of the       
chargino/neutralino mixing matrices $U$, $V$ and $N$
appearing in the
couplings: $
  U^{eff}=U+\Delta U^{(f)}   , \ 
  V^{eff}=V+\Delta V^{(f)}    ,  \   N^{eff}=N+\Delta N^{(f)}$. In this
  way we can \textit{decouple} the computation of the \textit{universal}
  (or \textit{super-oblique}~\cite{Katz:1998br}) corrections. These
  corrections contain the 
  non-decoupling logarithms from sfermion masses. 
\end{enumerate}
As an example of the \textit{universal} corrections we have computed
the electron-selectron contributions to the $\Delta U^{(f)}$ and $\Delta V^{(f)}$
matrices, assuming zero mixing angle in the
selectron sector ($\theta_e=0$), we have identified the leading terms in the
approximation $m_{\tilde e_i}, m_{\tilde \nu}\gg (\mw,\mi) \gg m_e$, and
 analytically cancelled the divergences and the  renormalization scale
 dependent terms;
finally, we have kept only the terms logarithmic in the {slepton}
masses. The result for $\Delta U^{(f)}$ reads as follows:
\begin{eqnarray}
\Delta\UChaio^{(f)}&=& \commonnumfactor\log\left(\frac{M^2_{\tilde e_L}}{\mxs}\right)\,\bigg[
 \frac{\UChaio^3}{6} - 
  \UChait \frac{\sqrt{2}\,\mw\,(M\,\cbta +
  \mu\,\sbta)}{3\,(M^2-\mu^2)\,\mchasomenysmchast^2}
\left(M^4 - M^2\,\mu^2 + \right.\nonumber\\
&&\left. + 3\,M^2\,\mws + 
     \mu^2\,\mws + \mwf + \mwf\,\cfbt 
    + (\mu^2-M^2)\,\mis + 
     4\,M\,\mu\,\mws\,\stbt\right)\,\bigg]\,\,,
\nonumber\\
\Delta\UChait^{(f)}&=&\commonnumfactor\log\left(\frac{M^2_{\tilde e_L}}{\mxs}\right)\,
\UChaio\,\frac{\mw\,(M\,\cbta + \mu\,\sbta)}
 {3\,\sqrt{2}\,(M^2-\mu^2)\,\mchasomenysmchast^2} \times \nonumber\\
  &\times&  \left((M^2-\mu^2)^2 
+ 4\,M^2\,\mws + 4\,\mu^2\,\mws + 2\,\mwf + 
   2\,\mwf\,\cfbt + 8\,M\,\mu\,\mws\,\stbt\right)\,\,,
\label{eq:logterms}
\end{eqnarray}
$M^2_{\tilde e_L}$ being the soft-SUSY-breaking mass of the
$(\tilde{e}_L,\tilde{\nu})$ doublet,
whereas $\mx$ is a SM mass.
In the on-shell scheme for the SM electroweak theory we define
parameters at very different scales, basically $\mx=\mw$ and
$\mx=m_e$.  These wide-ranging scales enter the structure of the
counterterms and so
must appear in eq.(\ref{eq:logterms}) too. As a result the leading log
in the various terms of this equation will vary accordingly. For
simplicity in the notation we have factorized $\log M^2_{\tilde
e_L}/\mxs$ as an overall factor. In some cases this factor can be very
big, $\log M^2_{\tilde e_L}/m_e^2$; it comes from the electron-selectron
contribution to the chargino-neutralino self-energies.  

\begin{figure}[tp]
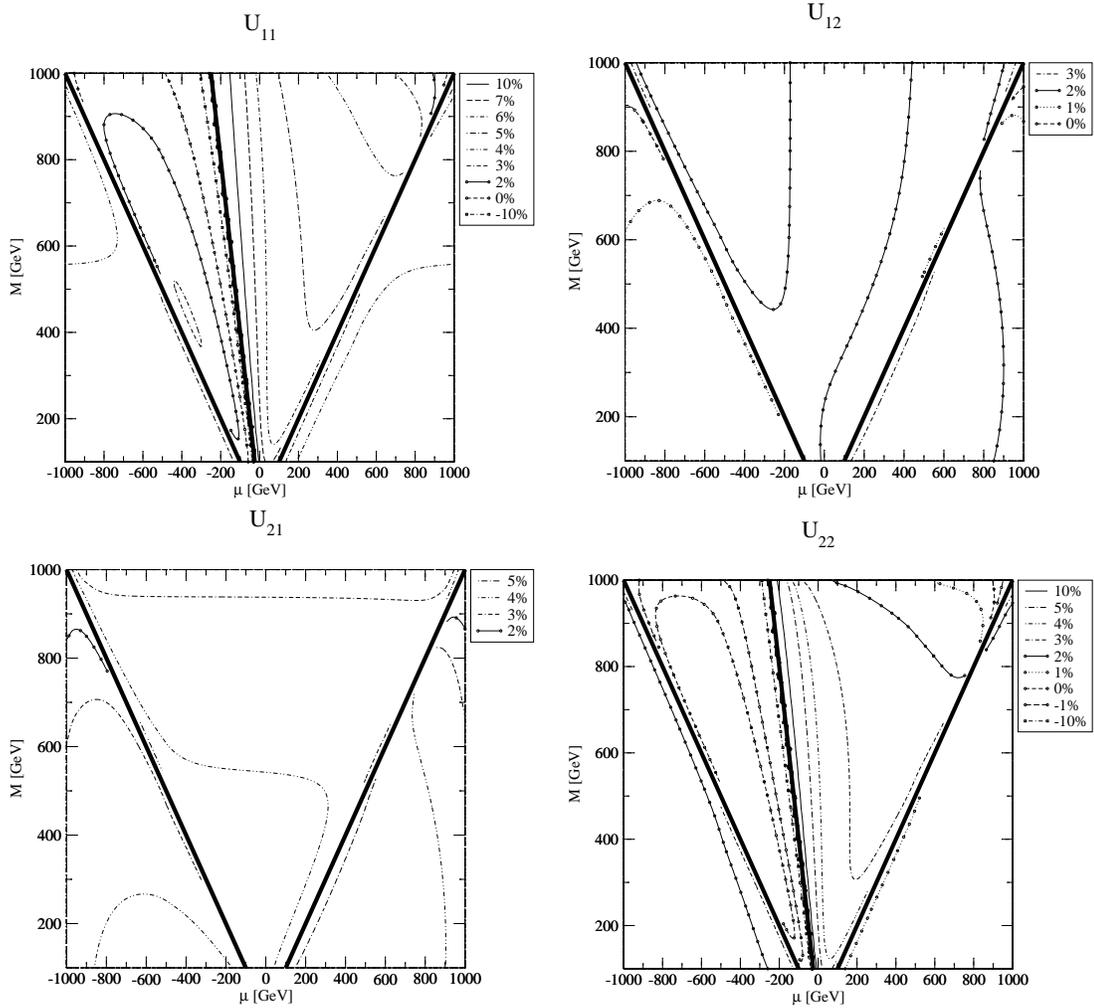

  \centering
\begin{tabular}{cc}
\resizebox{7cm}{!}{\includegraphics*{DUeff11.eps}}
&
\resizebox{7cm}{!}{\includegraphics*{DUeff12.eps}}
\\
\resizebox{7cm}{!}{\includegraphics*{DUeff21.eps}}
&
\resizebox{7cm}{!}{\includegraphics*{DUeff22.eps}}
\end{tabular}
  \caption{Relative correction to the effective chargino coupling matrix $\Delta U^{(f)}/U$
     in the $M-\mu$ plane, for $\tb=4$ and a
    sfermion spectrum around $1\TeV$~($m_{\tilde{l}_2}=m_{\tilde{d}_2}=m_{\tilde{u}_2}=1 \TeV \,,\,
 m_{\tilde{l}_1}=m_{\tilde{d}_1}=m_{\tilde{l}_2}+5\GeV\,,\,
\theta_l=\theta_q=\theta_b=0 \,,\, \theta_t=-\pi/5$).} 
  \label{fig:Ueff}
\end{figure}

In Fig.~\ref{fig:Ueff} we show the relative correction to the matrix
elements of $U$ for a sfermion spectrum around $1\TeV$. 
The thick black lines in Fig.~\ref{fig:Ueff} correspond
to {spurious} divergences in the relative corrections due to the
renormalization prescriptions.
Corrections as
large as $\pm10\%$ can only be found in the 
vicinity of these divergence lines. However, there exist large regions
of the $\mu-M$ plane  where the corrections are larger than $2\%$,
$3\%$, or even $4\%$.

\begin{figure}[tp]
  \begin{center}
    \begin{tabular}{cc}
\resizebox{7cm}{!}{\includegraphics{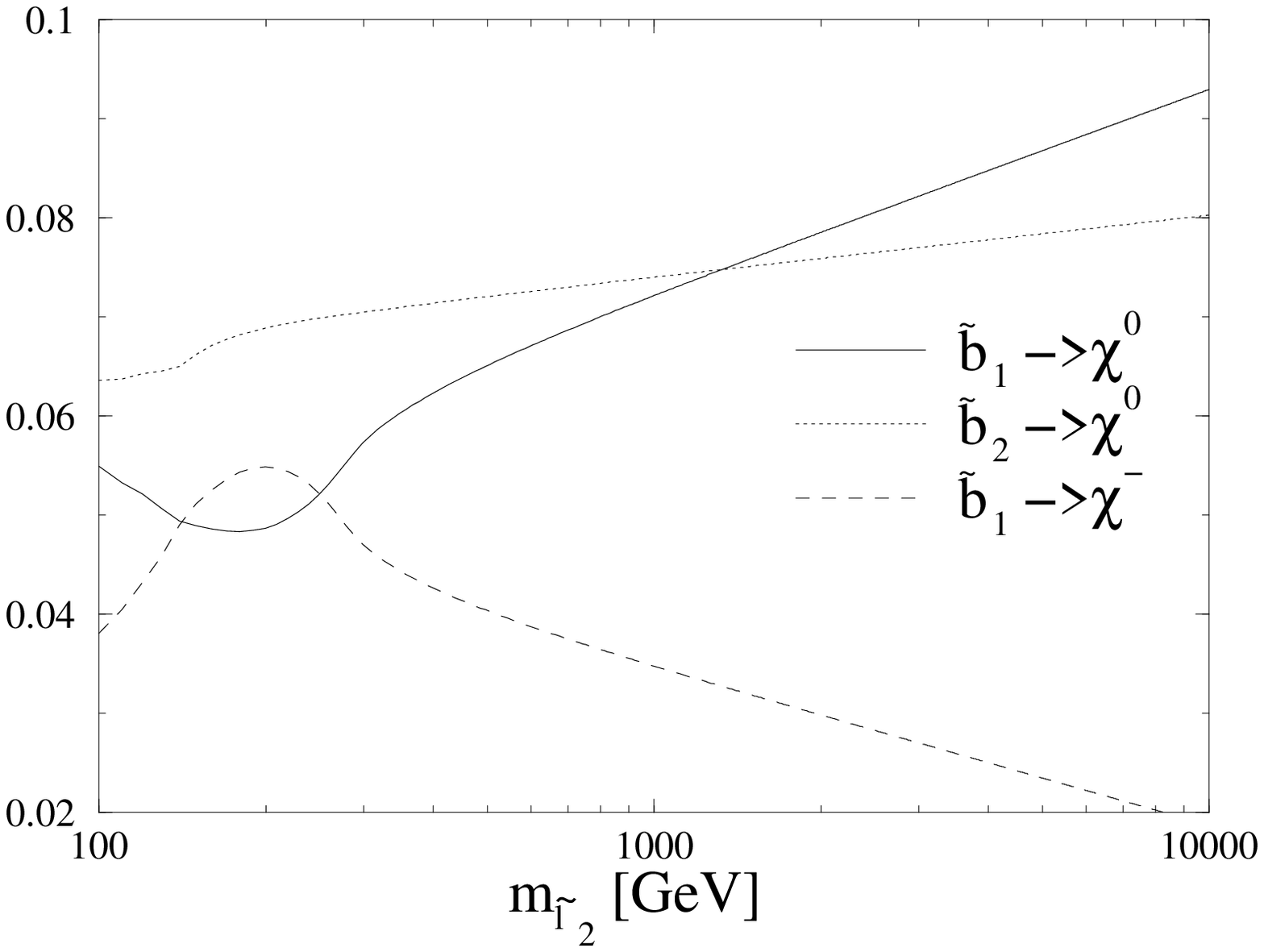}}&
\resizebox{7cm}{!}{\includegraphics{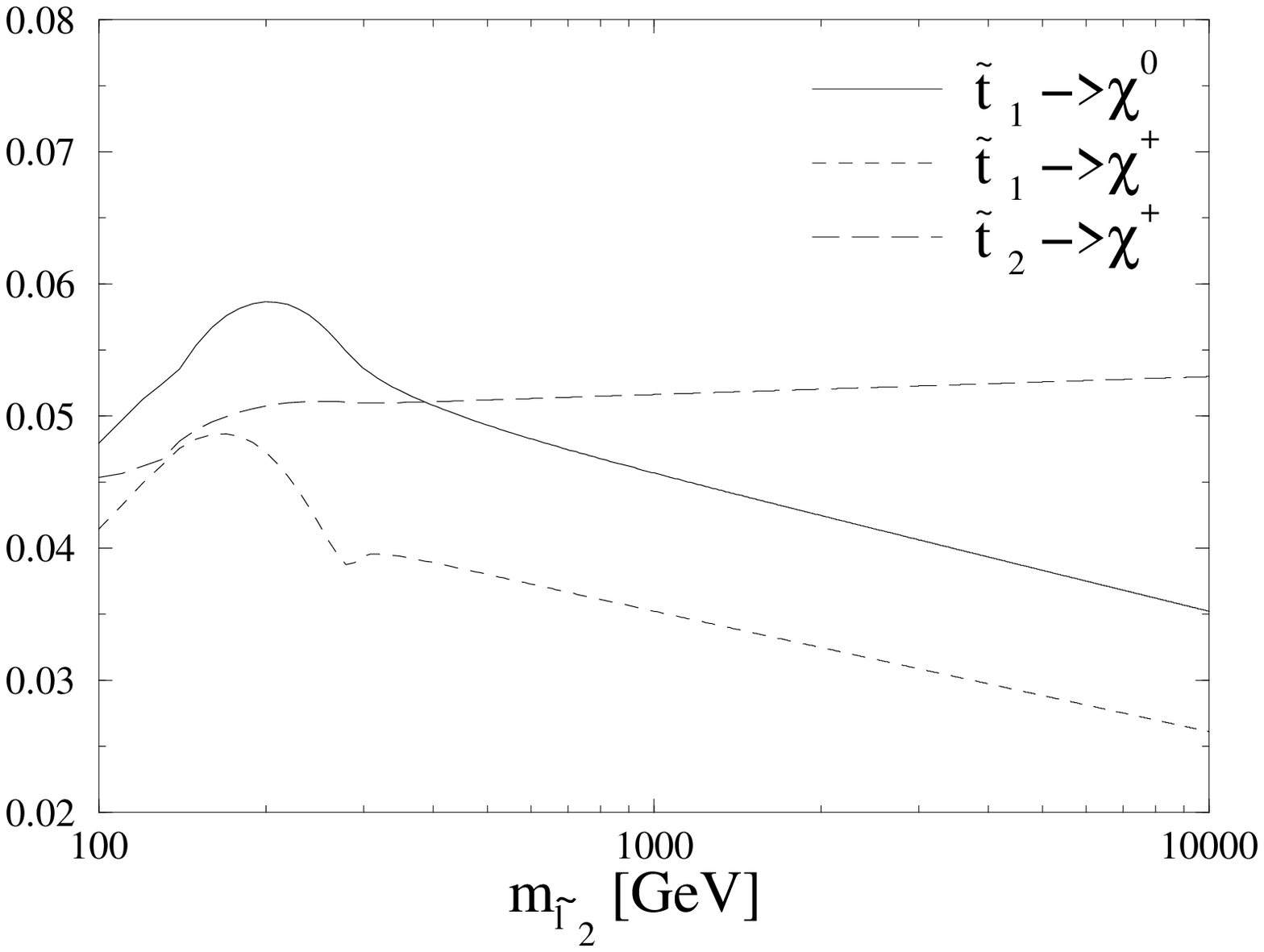}}
\\(a)&(b)
    \end{tabular}
  \end{center}
\caption{Universal relative corrections~(\ref{eq:totalcn}) to third generation squark partial decay widths as a
  function of a common slepton mass using the input parameter set~(\ref{eq:inputpars}).}  \label{fig:unisq}
\end{figure}

The effects of the universal corrections to the partial decay
widths~(\ref{eq:gammadef}) are shown in Fig.~\ref{fig:unisq} for top-
and bottom-squark decays as a function of a common slepton mass. 
Here (and in most of the 
discussion below) we show
the corrections to the total decay widths of sfermions into charginos and
neutralinos, that is
\begin{equation}
  \label{eq:totalcn}
\delta(\tilde{f_a}\to f'\chi)=\frac{ \sum_r 
    \left(\Gamma(\tilde{f_a}\to  f'\chi_r)-\Gamma^0(\tilde{f_a}\to  f'\chi_r)\right)}
  {\sum_r \Gamma^0(\tilde{f_a}\to  f'\chi_r)}\,\,, 
\end{equation}
with $\chi=\chi^\pm$ or $\chi=\neut$. 
We will not show results for processes {whose} branching ratio
{are} less that 10\% in all of the explored parameter space. The default
parameter set used is:
\begin{equation}
\begin{array}{l}
 \tb      =  4\,\,,\mt=175\GeV\,\,,
 \mb=5\GeV\,\,,
 \msbt=\msdt=\msut=\mselt=300\GeV\,\,,\\
 \msbo=\msdo=\mselo=\msbt+5\GeV\,\,,
\msut=290\GeV\,\,,
 \mstt=300\GeV\,\,,\\
 \osb=\osd=\osu=\osel=0\,\,,
 \ost=-\pi/5\,\,, 
 \mu      =  150\GeV\,\,,
 M       =  250\GeV\,\,,
 \mHp  =  120\GeV\,\,,
\end{array}
\label{eq:inputpars}
\end{equation}
The logarithmic behaviour from eq.~(\ref{eq:logterms}) is evident in this figure. 
{The logarithmic regime is attained already for slepton masses of
  order $1\TeV$.}
The universal corrections are seen to be
positive for all squark decays, ranging between $4\%$ and $7\%$ for
slepton masses below $1\TeV$.

Although above we have singled out the non-decoupling properties of
sfermions, we would like to stress that the whole spectrum shows
non-decoupling properties. By numerical analysis we have been able to
show the existence of logarithms of the gaugino mass parameters ($M/M_X$ and
$M'/M_X$), and the Higgs mass ($\mHp/M_X$). However, due to the
complicated mixing 
structure of the model, we were not able to derive simple analytic
expressions containing these non-decoupling logarithms. 
Note that in \textit{any} observable which includes the
fermion-sfermion-chargino/neutralino Yukawa couplings at leading order
we will have this 
kind of corrections, therefore the full MSSM spectrum must be taken into
account when computing radiative corrections, since otherwise one could
be missing large logarithmic contributions of the heavy masses.

\begin{figure}[tbp]
  \begin{center}
    \begin{tabular}{cc}
\resizebox{7cm}{!}{\includegraphics{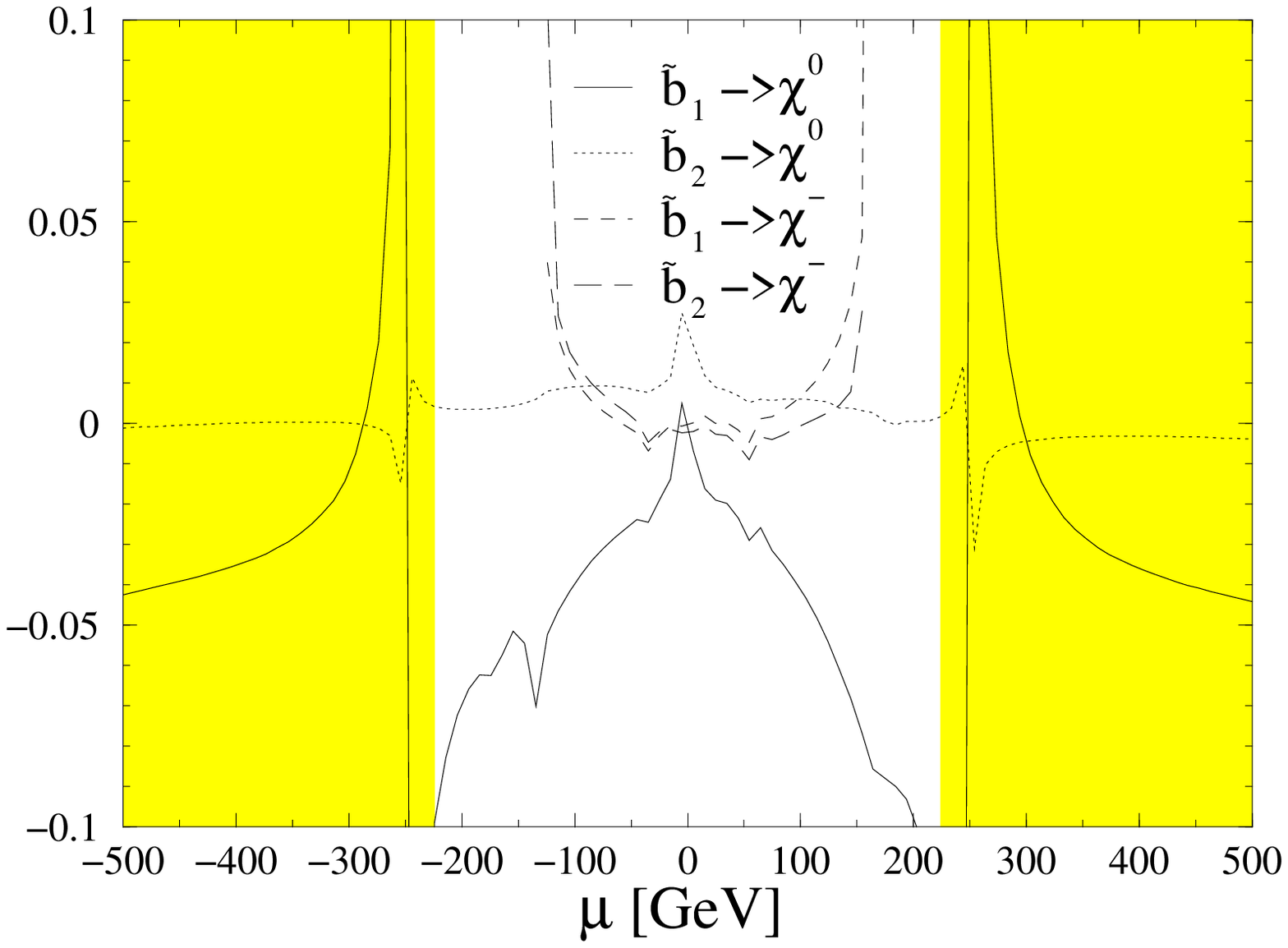}}&
\resizebox{7cm}{!}{\includegraphics{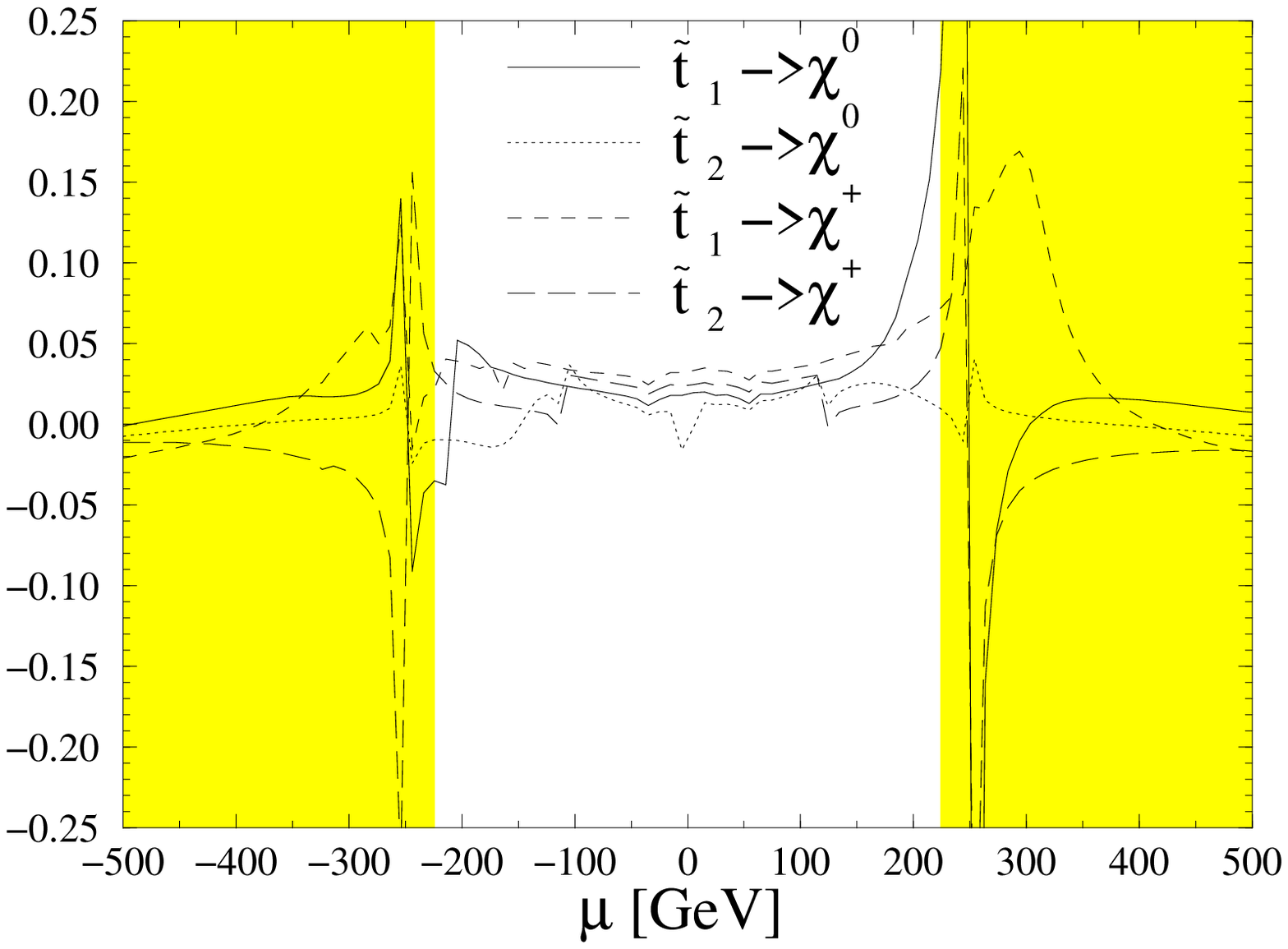}}
\\(a)&(b)
    \end{tabular}
  \end{center}
  \caption{Non-universal corrections to the partial decay width of
top- and bottom-squarks as a function of the higgsino mass parameter
    $\mu$. 
  {The shaded regions correspond to the violation of the
  condition~(\ref{eq:necessary}).}}\label{fig:muquark}
\end{figure}

As for the \textit{non-universal} part of the contributions, they show a
rich structure, as can be seen in Fig.~\ref{fig:muquark}. There we show
the evolution of the corrections as a function of the $\mu$ parameter
for top- and bottom-squark decays. 
A number of divergences are seen in the figure,
ones related to the mass renormalization framework (at $|\mu|=M$), and others
due to threshold singularities in the external wave function
renormalization constants. It is clear that the precise value of the
corrections is very much dependent on the correlation among the
different SUSY masses.

An important contribution to the corrections of third-generation
sfermion decays is the \textit{threshold correction} to the bottom-quark
($\tau$-lepton) Yukawa coupling 
($\Delta m_{\{b,\tau\}}$)~\cite{DmbTeo}. In the processes under
study~(\ref{eq:gammadef}) two kind of contributions appear: first, the
genuine corrections $\Delta m_{\{b,\tau\}}$ from SUSY loops in the
fermion self-energy; and second in the loops of sfermion self-energies
mixing different chiral states $\sfr_L \leftrightarrow \sfr_R$. This
kind of corrections grow with the sfermion mass splitting, the sfermion mixing
angle, and $\tb$. 

\begin{figure}
  \centering
\begin{tabular}{cc}
\resizebox{7cm}{!}{\includegraphics{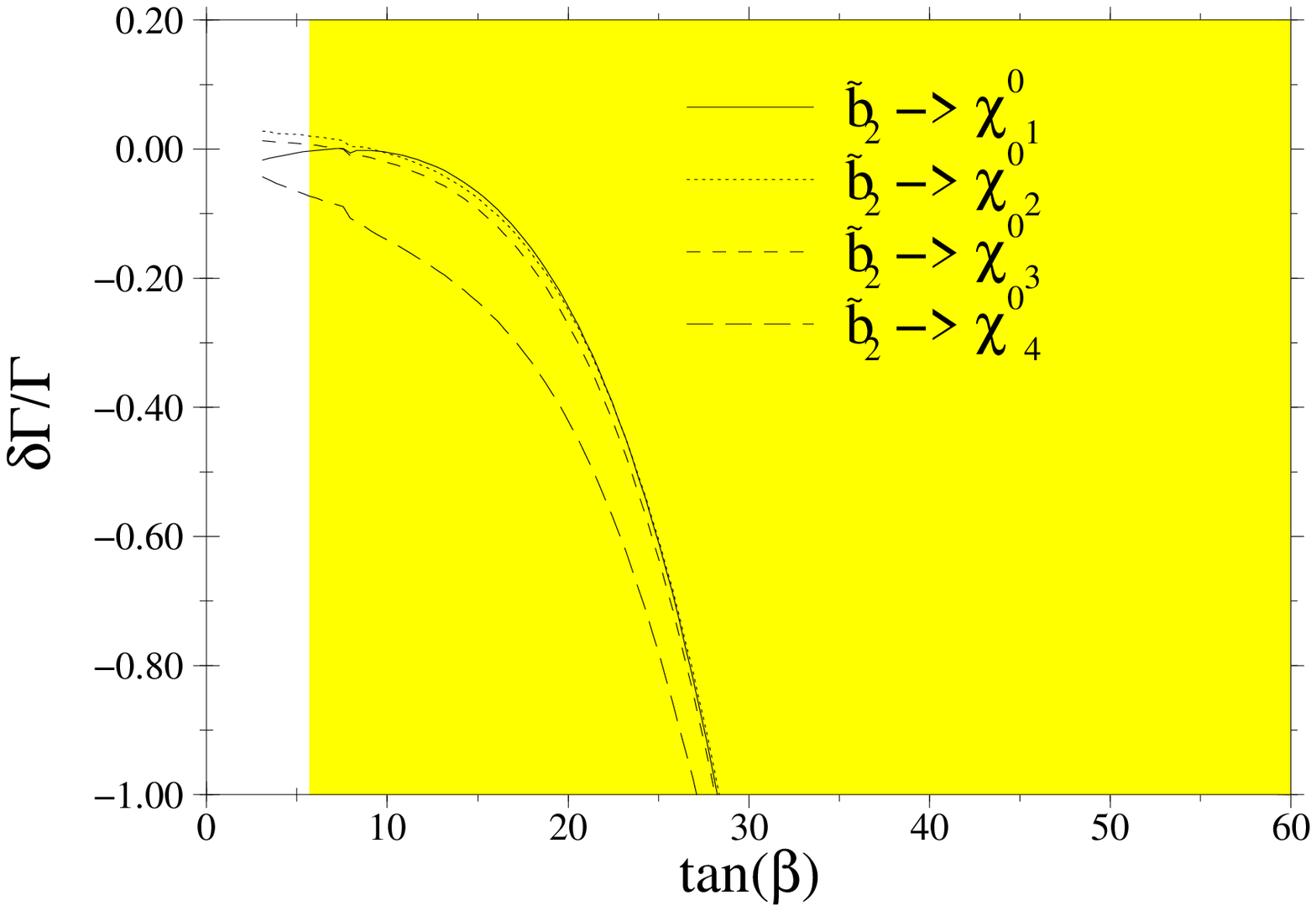}} &
\resizebox{7cm}{!}{\includegraphics{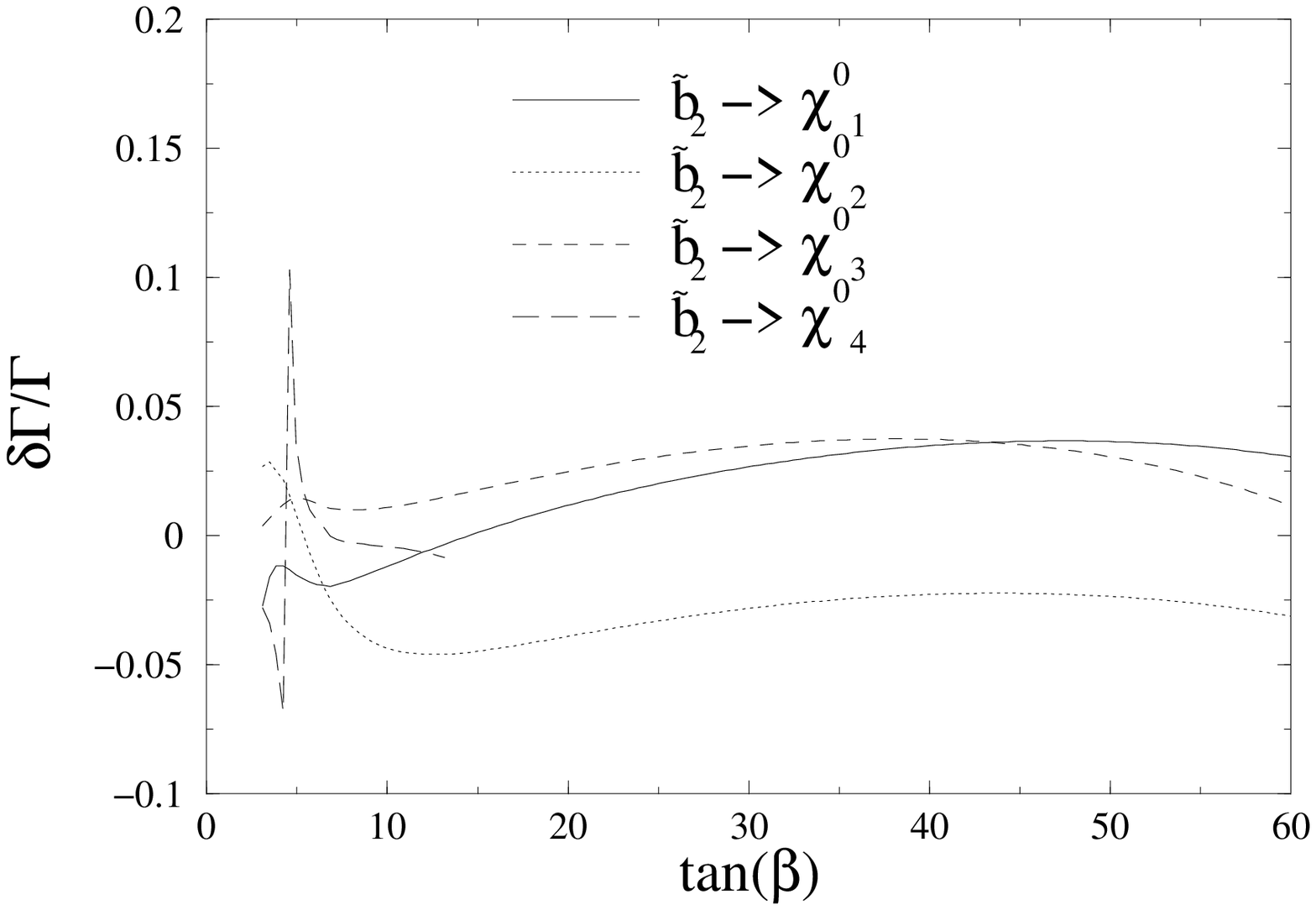}} \\
(a) & (b) 
\end{tabular}  
\caption{Non-universal relative corrections to the lightest
  bottom-squark partial decays 
  widths into  neutralinos as a function of
  $\tb$. \textbf{a)} Keeping fixed the splitting between the
  bottom-squarks $\msbo-\msbt=5\GeV$. \textbf{b)} Keeping $A_b=600\GeV,
  A_t=-78\GeV$.   {The shaded region corresponds to the violation of the
  condition~(\ref{eq:necessary}).}}  
\label{fig:sbottomcomp}
\end{figure}

A complementary set of corrections corresponds to the genuine
three-point vertex functions including Higgs bosons in the loops. These
contributions are proportional to the soft SUSY-breaking trilinear
couplings~(\ref{eq:Abt}), and therefore potentially large. Concretely,
if $\tb$ is large, and the bottom-squark mass splitting (or the mixing
angle) is small, the bottom-squark trilinear coupling grows with $\tb$
($A_b\simeq \mu\tb$), eventually inducing corrections larger than
$100\%$, spoiling the validity of perturbation theory. 
In Fig.~\ref{fig:sbottomcomp}a we show the evolution of the
corrections to the lightest bottom-squark decay into neutralinos as a
function of $\tb$ using the parameter 
set~(\ref{eq:inputpars}). We see the fast growing of the corrections,
reaching $-100\%$ at $\tb\simeq30$. Fortunately, applying the (necessary) 
restriction~(\ref{eq:necessary}) keeps the $A_q$ parameter small. In
Fig.~\ref{fig:sbottomcomp}b we show again the evolution of the
corrections as a function of $\tb$, but this time keeping a fixed value
for the trilinear couplings $A_b=600\GeV$, $A_t=-78\GeV$. The figure
shows that the corrections stay well below $10\%$ all over the $\tb$
range for this channel.

The complementarity between the $\Delta m_{\{b,\tau\}}$-like and the
$A_f$-like corrections is as follows: at large $\tb$, if the
bottom-squark mass splitting is large, there will be large corrections of
type $\Delta m_{\{b,\tau\}}$; on the other hand, if the bottom-squark
mass splitting is small, there will be large corrections of the type
$A_f$. Note that the QCD corrections contain $\Delta m_b$ terms but not
$A_f$ terms. When analyzing QCD corrections alone, one could choose a
small splitting, obtaining small corrections, however we have seen that
this is inconsistent, so one is forced to a large  $\Delta\mb^{QCD}$
contribution, which can reinforce (or screen) the negative corrections
from the standard running of the QCD coupling constant\footnote{Though
  it is not possible to separate between standard gluon corrections and
  gluino corrections, one can talk qualitatively about the
  contributions of the different sectors.}.

It is known that the electroweak corrections to any process grow as the
logarithm squared of the process energy scale due to the Sudakov
double-logs~\cite{sudakov}. We have observed this behaviour in the
process under study. 

At the end of the day, we want to analyze the branching ratios, which
are the true observables. For this analysis we have to add the QCD
corrections to the EW corrections. Due to the large value of the QCD
corrections, we made use of the enhanced resummed expression for
the bottom-quark Yukawa coupling~\cite{davidDmb}.
In Table~\ref{tab:brcorr} we show the tree-level and corrected branching
ratios for 
top- and bottom-squarks using the input parameter
set~(\ref{eq:inputpars}) and $\mg=500\GeV$. From inspection of
Table~\ref{tab:brcorr} we 
 see that the EW corrections can induce
 a change on the branching ratios of the leading decay channels of squarks
 comparable to the QCD corrections. Therefore both contributions must be
 taken into account on equal footing in the analysis of the phenomenology
 of sfermions.

 \begin{table}[tbp]
 \centering
   \begin{tabular}{|c|c|c|c|c||c|c|}
 \hline  
  & $\neut_1$ & $\neut_2$ & $\neut_3$ & $\neut_4$ & $\cplus_1$ & $\cplus_2$\\ \hline  
 $BR^{tree}(\tilde{t}_1 \to q\chi)$  & 0.169 & 0.249 & 0.145 & - & 0.159 & 0.278\\ \hline  
 $BR^{QCD}(\tilde{t}_1\to q\chi)$ & 0.164 & 0.257 & 0.144 & - & 0.099 & 0.335\\ \hline 
 $BR^{total}(\tilde{t}_1\to q\chi)$ & 0.177 & 0.242 & 0.143 & - & 0.122 & 0.316\\ \hline \hline 
 $BR^{tree}(\tilde{t}_2 \to q\chi)$ & 0.058 & - & - & - & 0.942 & -\\ \hline  
 $BR^{QCD}(\tilde{t}_2\to q\chi)$ & 0.063 & - & - & - & 0.937 & -\\ \hline 
 $BR^{total}(\tilde{t}_2\to q\chi)$ & 0.065 & - & - & - & 0.935 & -\\ \hline \hline 
 $BR^{tree}(\tilde{b}_1 \to q\chi)$  & 0.272 & 0.092 & 0.047 & 0.014 & 0.575 & -\\ \hline  
 $BR^{QCD}(\tilde{b}_1\to q\chi)$ & 0.308 & 0.104 & 0.031 & 0.018 & 0.538 & -\\ \hline 
 $BR^{total}(\tilde{b}_1\to q\chi)$ & 0.291 & 0.092 & 0.031 & 0.018 & 0.568 & -\\ \hline \hline 
 $BR^{tree}(\tilde{b}_2 \to q\chi)$ & 0.502 & 0.332 & 0.123 & - & 0.042 & -\\ \hline  
 $BR^{QCD}(\tilde{b}_2\to q\chi)$ & 0.541 & 0.386 & 0.054 & - & 0.019 & -\\ \hline 
 $BR^{total}(\tilde{b}_2\to q\chi)$ & 0.528 & 0.395 & 0.056 & - & 0.020 & -\\ \hline 
 \end{tabular}
 \caption{Tree-level and corrected branching ratios of top- and
 bottom-squark decays 
   into charginos and 
  neutralinos for the parameter set~(\ref{eq:inputpars}) and
   $\mg=500\GeV$.  Branching 
  ratios below $10^{-3}$ are not shown.} 
\label{tab:brcorr}
\end{table}

\vspace{0.2cm}

\noindent{Acknowledgments:}
The calculations have been done using the QCM cluster of the 
DFG Forschergruppe ``Quantenfeldtheorie, Computeralgebra und
Monte-Carlo Simulation''.
This collaboration is part of the network ``Physics at Colliders'' of the
European Union under contract HPRN-CT-2000-00149.
The work of J.G. has been partially supported by the 
European Union under contract No. HPMF-CT-1999-00150. 
The work of J.S.
has been supported in part by MECYT and FEDER under project FPA2001-3598.

\providecommand{\href}[2]{#2}

\end{document}